\documentclass[baaa]{baaa}

\usepackage[pdftex]{hyperref}
\usepackage{subfigure}
\usepackage{natbib}
\usepackage{helvet,soul}
\usepackage[font=small]{caption}

\begin{document}


\journalvol{60}
\journalyear{2018}
\journaleditors{P. Benaglia, A.C. Rovero, R. Gamen \& M. Lares}


\contriblanguage{0}


\contribtype{2}

\thematicarea{1}

\title{Propiedades fotométricas de las  galaxias enanas de bajo brillo
  superficial en la zona central del grupo Pegasus I}


\titlerunning{Propiedades  fotométricas  de  las galaxias  enanas  del
  grupo Pegasus I}


\author{N.\,Gonz\'alez\inst{1,2},
        S.\,A.\,Cellone\inst{1,3,4},
        A.\,Smith\,Castelli\inst{1,2,3},
        F.\,Faifer\inst{1,2,3},
	C.\,Escudero\inst{1,2,3}
       }
\authorrunning{González et al.}


\contact{ngonzalez@fcaglp.unlp.edu.com.ar}

\institute{Facultad de Ciencias Astron\'omicas y Geof\'isicas,  
        UNLP, La Plata, Argentina.
        \and  
        Instituto de Astrof\'isica de La  Plata (CCT La Plata, CONICET
        - UNLP), La Plata, Argentina.
        \and 
        Consejo   Nacional   de    Investigaciones   Cient\'ificas   y
        T\'ecnicas, Argentina.
        \and
        CASLEO, San Juan, Argentina.
}


\resumen{
En este trabajo presentamos los  resultados preliminares de un estudio
fotométrico   de    objetos   de    bajo   brillo    superficial   con
$\mu_{g'}~\gtrsim~25$~mag/arcsec$^2$,   presentes  en   varios  campos
obtenidos con GEMINI-GMOS en la región central del grupo de Pegasus I.
Encontramos que sus características  fotométricas resultan similares a
las denominadas  galaxias enanas esferoidales  o de ultra  bajo brillo
superficial.
}
\abstract{
Here we  show the preliminary  results of  a photometric study  of low
surface brightness  objects with $\mu_{g'}~\gtrsim~25$~mag/arcsec$^2$.
These objects are present in several fields obtained with GEMINI-GMOS,
in the  central region of  the Pegasus I  group.  We found  that their
photometric  characteristics  are  similar   to  the  so-called  dwarf
spheroidal galaxies or to the ultra low surface brightness galaxies.
}
%
%
\maketitle
\section{Introducción}
\label{S_intro}
A pesar de  los numerosos estudios de galaxias de  baja luminosidad en
diversos medioambientes, aún no hay  un consenso generalizado sobre su
escenario  de formación.   Debido a  esto, es  de gran  importancia la
detección y  análisis de dicho tipo  de objetos.  Se espera  que tales
estudios proporcionen  condiciones de contorno a  los modelos actuales
sobre la formación y evolución de las galaxias, as\'i como tambi\'en a
los de  formaci\'on de la estructura  a gran escala del  Universo.  En
este contexto, presentamos  el estudio fotométrico de  la población de
galaxias de  tipo temprano y  de bajo  brillo superficial (LSB)  en el
grupo Pegasus\,I.

Pegasus\,I es un pequeño grupo de galaxias, localizado a una distancia
de  50~Mpc  \citep{2001ApJ...546..681T}, y  dominado por  dos galaxias
elípticas masivas:  NGC\,7626 y  NGC\,7619.  Este grupo  representa un
ambiente  de especial  interés porque  los estudios  de la  emisión en
rayos-X del gas  caliente intragrupo indican que  Pegasus I constituye
la fusión en desarrollo de los  dos subgrupos asociados a cada galaxia
dominante \citep{2009ApJ...696.1431R}.

A pesar  de que Pegasus\,I  fue estudiado  ampliamente en cuanto  a su
poblaci\'on          de         galaxias          tardías         {\bf
  \citep[p.ej.][]{2007AJ....133.1104L,2008MNRAS.389..341M}},        su
población de  galaxias de  tipo temprano se  encuentra pr\'acticamente
inexplorada.

En  este contexto,  Pegasus\,I representa  un excelente  medioambiente
para el estudio de  galaxias LSB y de los procesos  que se cree juegan
un papel en  su formación y evolución.  En este  trabajo mostramos los
resultados   preliminares   del    estudio   fotométrico   (magnitudes
integradas,        colores,         ajustes        de        perfiles,
$\langle\mu_{\mathrm{eff}}\rangle$,  etc.)    de  ocho   candidatas  a
galaxias       de       tipo       temprano      y       LSB       con
$\mu_{g'}~\gtrsim~25$~mag/arcsec$^2$, presentes en cinco campos de los
alrededores de las galaxias NGC\,7626 y NGC\,7619.
\section{Datos fotométricos}
\begin{figure}[!t]
  \centering
  \includegraphics[width=0.49\textwidth]{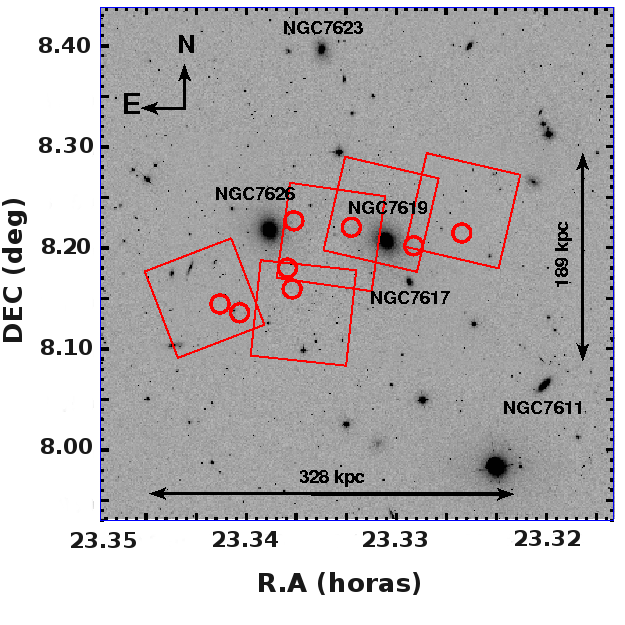}
  \caption{Mosaico de  $30.5\times~30.5$~arcmin  en el filtro  $r'$ de
    SDSS DR12 mostrando la región central del grupo de Pegasus\,I. Los
    marcos  rojos  corresponden  a  los campos  de  GEMINI-GMOS.   Los
    circulos rojos indican la ubicación de las candidatas LSB.
}
  \label{sdss_campos}
\end{figure}

\begin{figure}[!t]
  \centering
  \includegraphics[width=0.24\textwidth]{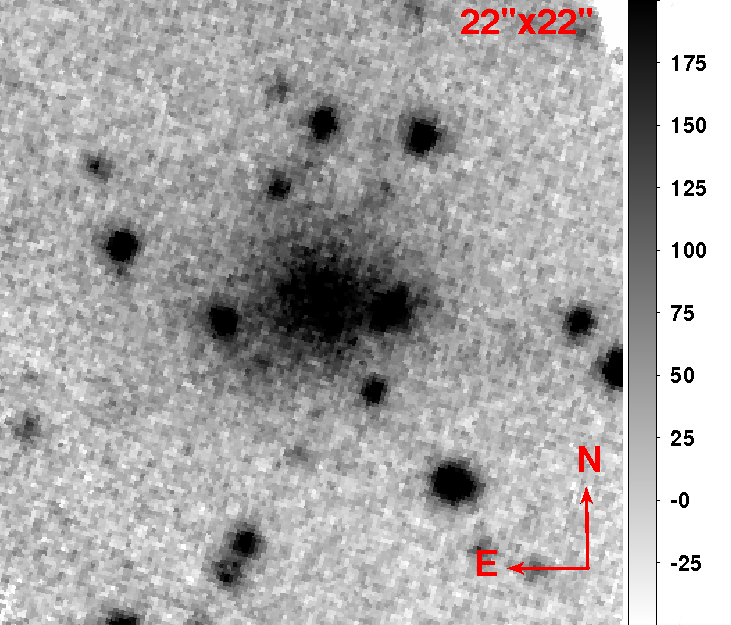}
  \includegraphics[width=0.24\textwidth]{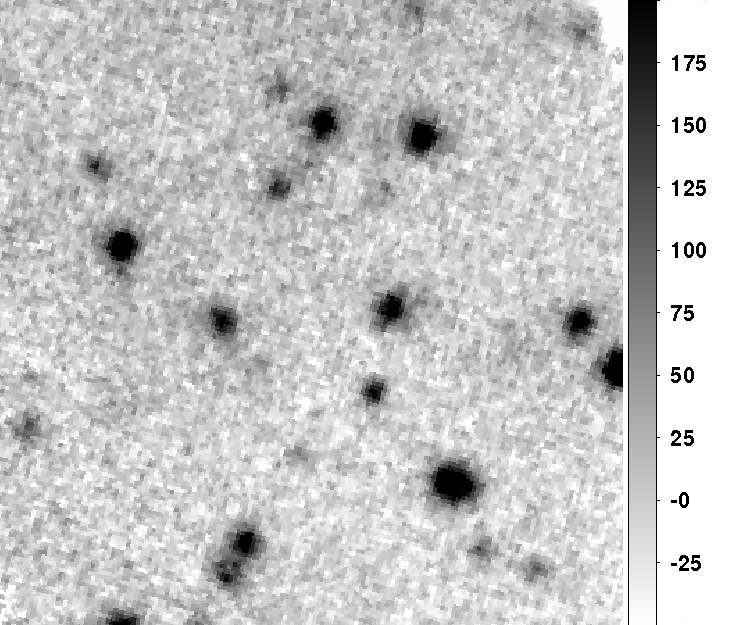}
  \includegraphics[width=0.499\textwidth]{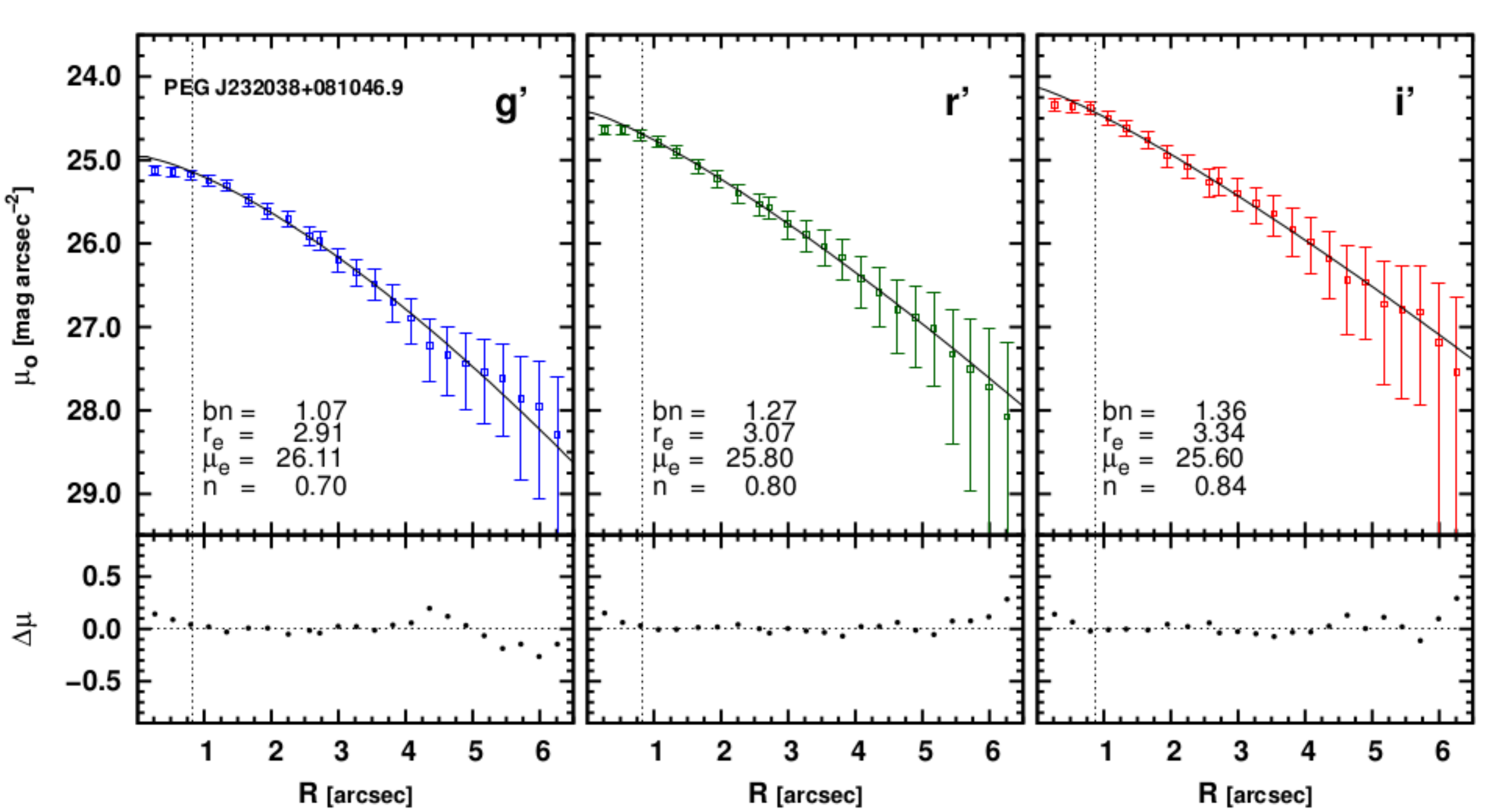}
  \caption{Paneles superiores: imagen de una de las galaxias LSB de la
    muestra en  el filtro $r'$  de GEMINI-GMOS; imagen residual  en el
    filtro  $r'$ obtenida  de  la  resta del  modelo  creado por  {\it
      ellipse} y {\it bmodel}.  Paneles  inferiores: ajustes de la ley
    Sérsic a los  perfiles de brillo $g'$, $r'$ e  $i'$ de la galaxia.
    Los residuos $\Delta\mu~=~\mu~(\mathrm{obs})~-~\mu~(\mathrm{fit})$
    se muestran en el panel inferior.  Las líneas verticales punteadas
    indican la región interna de  los perfiles excluidos para realizar
    los ajustes que  se corresponden al {\it seeing}  de las imágenes.
    Todos   los   perfiles   están    corregidos   por   extinción   y
    enrojecimiento.}
  \label{perfiles}
\end{figure}

Este  trabajo está  basado en  cinco campos  profundos tomados  en los
filtros $g'$,  $r'$ e  $i'$ \citep{1996AJ....111.1748F},  empleando la
cámara GMOS  de Gemini Norte  (Programa GN-2008B-Q-14, PI:  F. Faifer;
Programa   GN-2012A-Q-55,    PI:   A.    Smith    Castelli;   Programa
GN-2014B-Q-17,  PI:   F.   Faifer;  Programa  GN-2015B-Q-13,   PI:  N.
González).  Estas imágenes  cubren el entorno inmediato  a NGC\,7626 y
NCG\,7619, y se utilizaron para obtener los parámetros fotométricos de
ocho candidatas a galaxias LSB.  La Fig.~\ref{sdss_campos} muestra la
orientación  de los  diferentes campos  analizados y  la ubicación  de
estos objetos.   Por su parte,  en el  panel superior izquierdo  de la
Fig.~\ref{perfiles} , se muestra la  imagen de una de las candidatas a
galaxia LSB en el filtro $r'$.

Dado que no  se cuenta con datos espectroscópicos de  las candidatas a
galaxias LSB, la  identificación de las mismas se realizó  a partir de
una  inspección  visual  en  las  imágenes  GEMINI-GMOS.   Para  ello,
utilizando las tareas {\it ellipse} y {\it bmodel} de IRAF, primero se
procedió a  modelar las  distribuciones de  brillo superficial  de las
galaxias  elípticas NGC\,7626  y  NGC\,7619 y  sus respectivos  halos,
incluyendo  varios  objetos  extendidos.  Luego,  se  restaron  dichos
modelos con el fin de poder identificar y medir los perfiles de brillo
de las candidatas  a galaxias LSB.  Suponiendo que las  mismas están a
la  distancia  de  NGC\,7626,  y  adoptando  el  módulo  de  distancia
$m~-~M~=~33.67$~mag  \citep{2007ApJ...668..130C},  se  obtuvieron  sus
magnitudes  y colores  integrados,  con los  que  se construyeron  los
diagramas              color-magnitud             (DCM)              y
$\langle\mu_{\mathrm{eff}}\rangle$-luminosidad.

En la Fig.~\ref{perfiles} se presentan, en los paneles inferiores, los
perfiles de brillo superficial en los filtros $g'$, $r'$ e $i'$ de una
de las  galaxias LSB analizadas.  En  este caso, se puede  observar el
bajo brillo superficial que presenta dicho objeto. 

Se  realizaron ajustes  de Sérsic  \citep{1968adga.book.....S} en  los
filtros mencionados,  obteniendo los valores  de los índices  $n$ para
todas las candidatas. Dicho ajuste está dado por:
\begin{equation}
\mu (r)=\mu_{0}+1.0857~\left(\dfrac{R}{R_{0}} \right)^{\dfrac{1}{n}}\textrm{.}
\end{equation}
donde  $\mu_0$  es  el  brillo  superficial  central  y  $R_0$  es  un
par\'ametro de escala.
\section{Resultados preliminares}
Con  el  objetivo  de  obtener  una  clasificación  morfológica  y  de
pertenencia de las  candidatas a galaxias LSB al  grupo de Pegasus\,I,
comparamos  con diferentes  muestras  de tipo  temprano reportadas  en
varios trabajos.

\begin{figure}[!t]
  \centering
  \includegraphics[width=0.5\textwidth]{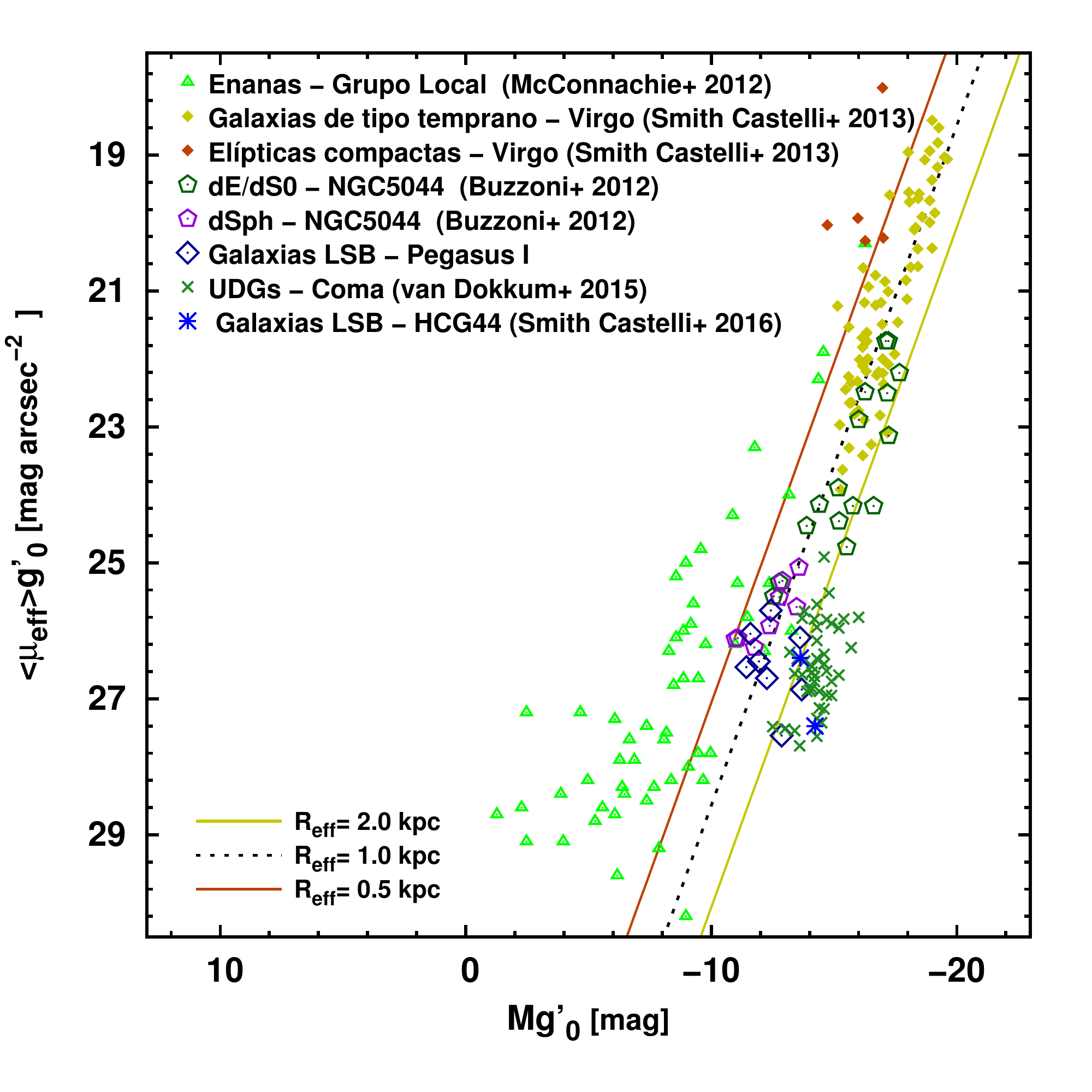}
  \caption{Diagrama $\langle\mu_{\mathrm{eff}}\rangle$ -luminosidad de
    las galaxias de  tipo temprano en la región central  del cúmulo de
    Virgo \citep{2013ApJ...772...68S},  mostrando la ubicación  de las
    galaxias LSB presentadas  en este trabajo, suponiendo  que están a
    la distancia de NGC\,7626.   Incluimos diferentes muestras de dSph
    y  galaxias   ultra-difusas  repor\-tadas   en  los   trabajos  de
    \citet{2012MNRAS.420.3427B},          \citet{2012AJ....144....4M},
    \citet{2015ApJ...798L..45V} y \citet{2016A&A...596A..23S}.
}
  \label{mueff}
\end{figure}

{\bf  La  Fig.~\ref{mueff}  muestra  la  relación  entre  el  brillo
  superficial efectivo  y la  luminosidad en  el filtro  $g'$. Podemos
  observar,} en el extremo brillante,  una muestra de galaxias de tipo
temprano en el cúmulo de Virgo \citep{2013ApJ...772...68S}. {\bf Estas
  galaxias tienden  a ubicarse en  el rango  de radio efectivo  $0.5 <
  R_{\mathrm{eff}}  <  2.0$~kpc.}   Sin  embargo,  las  galaxias  {\bf
  enanas} pertenecientes al Grupo Local (LG) tienden a ubicarse en una
zona  de radios  efectivos  más pequeños  \citep{2012AJ....144....4M}.
Por    otro     {\bf    lado,     las    galaxias     ultra    difusas
  \citep[UDGs,][]{2015ApJ...798L..45V}              y              LSB
  \citep{2016A&A...596A..23S} detectadas fuera  del LG muestran radios
  efectivos más grandes.}

A partir de  esto, podemos decir que parece existir  una distinción en
los tamaños:  los objetos en grupos  y cúmulos cercanos tienden  a ser
más  pequeños  que  los  reportados en  entornos  más  distantes.   En
consecuencia,  analizamos las  candidatas  a galaxias  LSB de  nuestra
muestra, que presentan similitudes con las denominadas galaxias enanas
esferoidales {\bf (dSph) y las UDGs}. 

\begin{figure}[!t]
  \centering
  \includegraphics[width=0.48\textwidth]{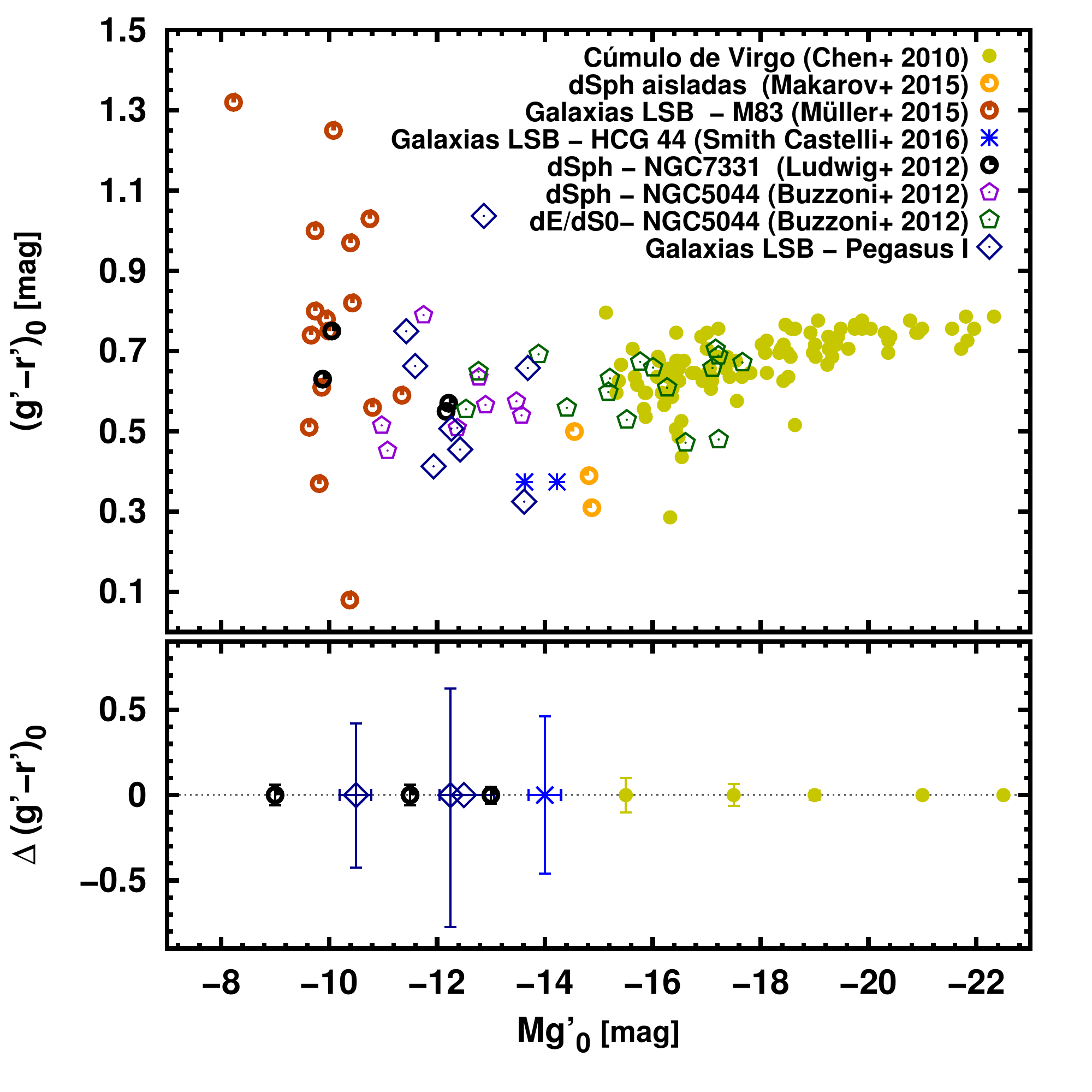}
  \caption{DCM de las  galaxias de tipo temprano en  la región central
    del  cúmulo  de  Virgo \citep{2010ApJS..191....1C},  mostrando  la
    ubicación  de  las  galaxias   LSB  presentadas  en  este  trabajo
    adoptando  que  están  a  la  distancia  de  NGC\,7626.  Incluimos
    diferentes muestras de dSph  y galaxias ultra-difusas repor\-tadas
    en        la        literatura        \citep{2012MNRAS.420.3427B},
    \citep{2015A&A...581A..82M},     \citep{2015A&A...583A..79M}     y
    \citep{2016A&A...596A..23S}.  En el panel inferior se muestran los
    errores medios  $\Delta (g'-r')_0$, tomados en  rangos de $\Delta$
    M$_{g'}=1$~mag.
  }
  \label{DCM}
\end{figure}

En la  Fig.~\ref{DCM} se presenta  el DCM  de los resultados  para las
galaxias de tipo de temprano obtenidos por \citep{2010ApJS..191....1C}
y  otros trabajos  que reportan  galaxias  LSB.  Se  destaca la  clara
relación para las galaxias de tipo  temprano del cúmulo de Virgo, y se
puede observar que algunas galaxias  LSB siguen dicha relación, aunque
algunas  otras se  apartan  tanto  hacia colores  más  rojos como  más
azules.  Por  lo tanto, en  términos de  colores, las galaxias  LSB no
parecen ser una clase homogénea de objetos.

\begin{figure}[!t]
  \centering
  \includegraphics[width=0.49\textwidth]{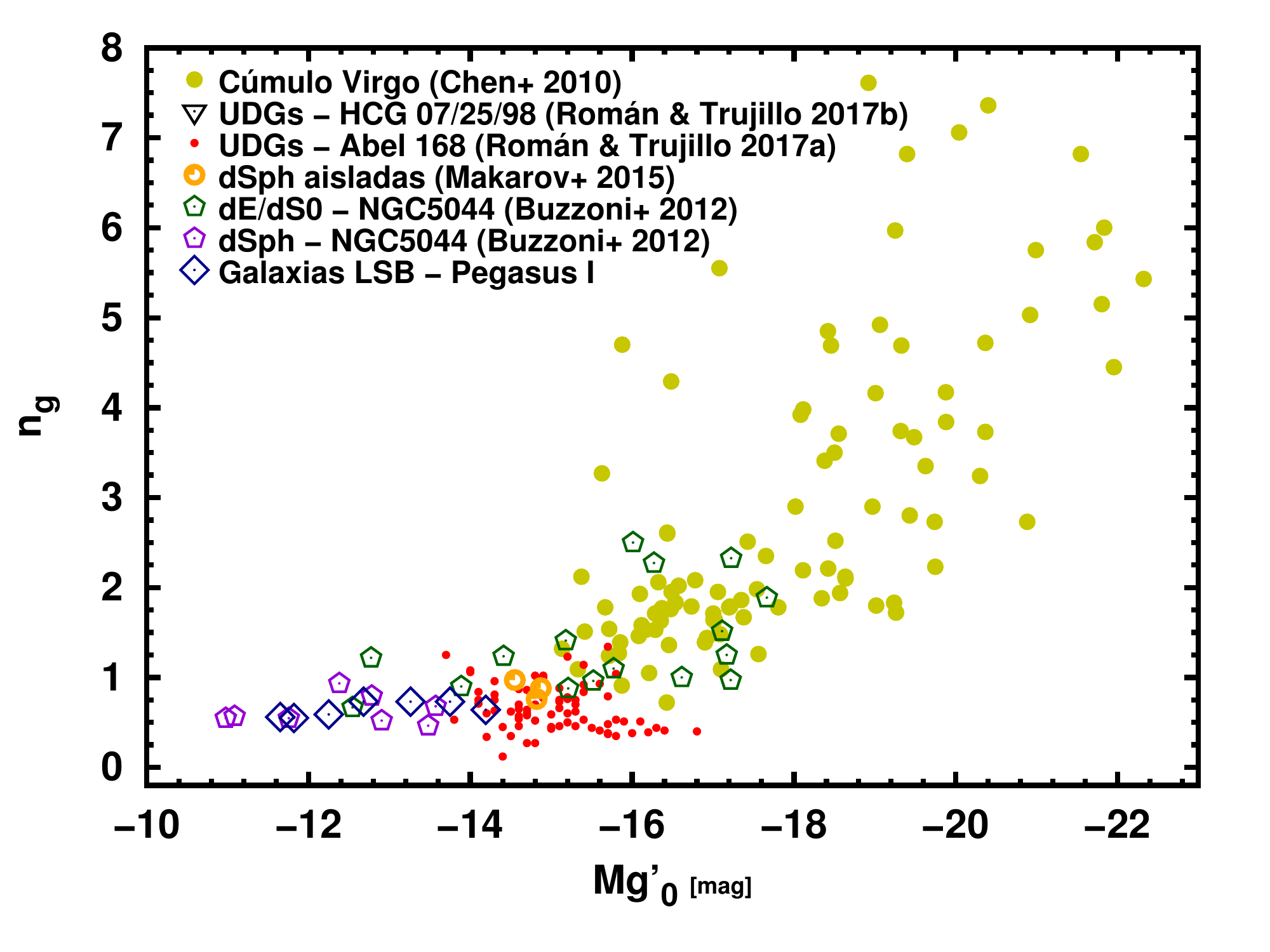}
  \caption{Relación  entre  el  parámetro  de  Sérsic  $n$  contra  la
    magnitud absoluta en  el filtro $g'$ para  nuestras candidatas LSB
    (rombos azules) adoptando  que están a la  distancia de NGC\,7626.
    Incluimos  diferentes  muestras  de   galaxias  de  tipo  temprano
    reportadas   en  los   trabajos  de   \citet{2010ApJS..191....1C},
    \citet{2012MNRAS.420.3427B},          \citet{2015A&A...581A..82M},
    \citet{2017MNRAS.468..703R} y \citet{2017MNRAS.468.4039R}.
}
  \label{sersic}
\end{figure}

En la Fig.~\ref{sersic} se presenta  la relación entre el parámetro de
Sérsic $n$ contra la magnitud absoluta  en el filtro $g'$, donde vemos
que existe una  relación entre $n$ y $M_{g'}$. Dicha  relación está de
acuerdo  con lo  encontrado  por distintos  autores,  entre ellos  los
trabajos       de        {\bf       \cite{1994ApJS...93..397C}       y
  \cite{2012MNRAS.420.3427B}} que  establecen que el índice  de Sérsic
debe ser $n$~$\lesssim$~1 para galaxias LSB.
\section{Trabajo a futuro}
%

En lo inmediato,  serán incorporados más datos similares a  los de las
candidatas a galaxias LSB de Pegasus\,I reportados en la literatura en
otros medioambientes, completando así  el análisis detallado  de estos
objetos presentado en Smith Castelli et~al. (en preparación~2018).
           
En  esta  dirección,  como  trabajo a  futuro  extenderemos  el  mismo
análisis  a las  candidatas a  enanas  típicas ya  localizadas en  los
campos observados  del grupo de  Pagasus I. Asimismo,  realizaremos un
trabajo  similar  en  cuatro   campos  observados  del  supergrupo  de
Eridanus. Resulta interesante realizar un estudio comparativo de ambos
grupos de galaxias de baja luminosidad. 
%







\bibliographystyle{baaa}
\small
\bibliography{biblio}

\end{document}